\documentclass[letterpaper,10pt]{article}
\usepackage{osameet2}
\DeclareGraphicsExtensions{.eps}
\usepackage{amsmath,amssymb}
\usepackage{subfig}
\usepackage[dvips,colorlinks=true,bookmarks=false,citecolor=blue,urlcolor=blue]{hyperref} 

\begin{document}

\title{Study of noise impact on nonlinear frequency division multiplexing}

\author{Wen Qi Zhang$^1$, Qun Zhang$^2$, Terence H. Chan$^2$ and Shahraam Afshar V.$^1$}

\address{1. Institute for Photonics and Advanced Sensing (IPAS), The University of Adelaide, 5005, Australia\\
2. : Institute for Telecommunications Research, University of South Australia, 5095, Australia}

\begin{abstract}
In this work, how the noise contaminates signals propagating in an optical fibre is discussed in the nonlinear spectral domain after applying the nonlinear Fourier transform. Simulation results about how the soliton parameters in the nonlinear spectral domain perturb is shown.
\end{abstract}


\section{Introduction}
The noisy signal propagation through an optical fiber is governed by the following normalized stochastic nonlinear Schrodinger equation (SNLSE)~\cite{Yousefi1_2014}
\begin{equation}
    jq_{z}(t,z)=q_{tt}(t,z)+2|q(t,z)|^{2}q(t,z)+j\epsilon G(t,z),
    \label{eq1}
\end{equation}
where $j=\sqrt{-1}$, and $q(t,z)$ is the electrical field propagation in the optical fiber. The term $G(t,z)$ is a standard circularly symmetric complex white Gaussian noise, and $\epsilon$ is a small scalar. Because of the fiber nonlinearity and chromatic dispersion effect, signals from different users interfere with each other in multiuser communications. Recently, Yousefi and Kschischang~\cite{Yousefi1_2014, Yousefi2_2014, Yousefi3_2014} propose a new transmission scheme, called nonlinear frequency division multiplexing (NFDM), with the help of the nonlinear Fourier transform, in which the time domain nonlinear channel can be diagonalized in the (nonlinear) spectral domain (composed by discrete spectrum and continuous spectrum) into several linear scalar multiplicative channels that are independent of each other. Hence, interference free multiuser communication can be achieved at least in the noise free case, which provides a potential of increasing the data rate of communication.

However, modulating continuous spectrum is complicated because a set of integration equations need to be solved. As a result, $N$-solitons are of interest in optical fiber communications, which does not have continuous spectrum, and only have $N$ discrete eigenvalues on the upper half complex plane. Zhang et. al~\cite{Zhang_2014} design a special sub-class of $N$-solitons, namely spatially periodic signals $q(t,z)$, as possible input signals for the NFDM, which maintain their shape periodically over $z$. In this paper, we restrict our input to be (fundamental) solitons, where $N=1$. Although the time domain behavior of solitons with noise is known, the spectral domain analysis is lacking. Our motivation is to study how the noise contaminates the solitons in the spectral domain. Specifically, how are the eigenvalue and the spectral amplitude perturbed due to noise. This work is of importance because the properties of the noise need to be understood in order to study the fundamental limit of reliably communication~\cite{Cover_2006}.

\section{Numerical Method}
We use a split-step nonlinear pulse propagation solver (NPPFS)~\cite{NPPS_2014} for solving the SNLSE. Uniformly distributed pseudo random numbers were generated on a computer and then transformed into Gaussian noises. The input of the simulation is an ideal fundamental soliton which follows the form of Eq.(\ref{eq1}):
\begin{equation}
    q(t,z) = 2 \beta(z) sech \left\{ 2 \beta(z) \left[ t - \frac{1}{2\beta(z)} ln \frac{\left|Q(z)\right|}{2\beta(z)} \right]\right\} e^{-2j\alpha(z)t-j\left\{arg\left[Q(z)\right]+\frac{\pi}{2}\right\}},
    \label{eq2}
\end{equation}
where $\alpha$ and $\beta$ are the real and imaginary part of the eigenvalue and $Q$ is the spectral amplitude. We chose $\alpha(0)=0$, $\beta(0)=0.5$, $|Q(0)|=1$ and $argQ(0)=\pi/2$ for the initial values.

Nonlinear curve fittings were applied on the amplitudes and the phases of the output pulses for variables $\alpha$, $\beta$, $|Q|$ and $arg Q$. Due to the existence of background noise in the FFT window greatly influences the fitting results, only the range of data where $|q|>0.5$ where used. Same calculations were repeated 100 times for each $\epsilon$.

\section{Results and Discussions}
\begin{figure}[h!]
    \centering
    \includegraphics[width=16cm]{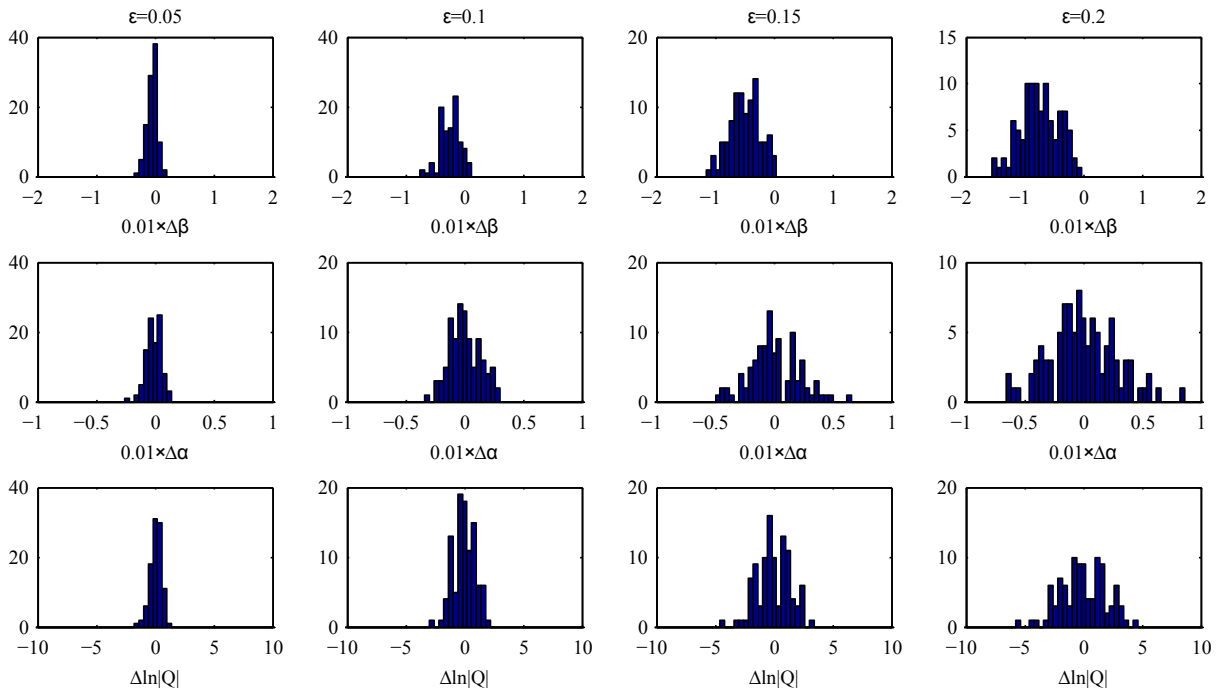}
    \caption{The histogram of the real (middle row) and imaginary (top row) part of the eigenvalues and the nonlinear spectral amplitude (bottom row) for different noise levels.}
    \label{fig1}
\end{figure}

The results of the simulations are shown in Fig.\ref{fig1}. The existence of the noise perturbs both real ($\alpha$) and imaginary ($\beta$) part of the eigenvalues, which translates into the changes in the output solitons' amplitudes and group velocities. As the noise increases, the fluctuations in both real and imaginary part of the eigenvalues increase. However, the overall changes are relatively small considering $\epsilon=0.2$ corresponding to approximately 10\% noise in the output pulse power. Furthermore, when the noise added is small ($\epsilon=0.05$), the fluctuation of the real and imaginary part of the eigenvalues are more or less evenly distributed around the mean values. As the noise increases, the distribution of the imaginary part of the eigenvalues shifts towards negative side, which corresponding to a decrease in soliton amplitude. The reason for the decrease can be understood as the power transfer from the soliton to noise through nonlinear interactions as such cross-phase modulation and modulation instability~\cite{agrawal_nonlinear_2001}.

\section{Conclusion}
In this paper, we numerically study the noise contaminates the solitons in the nonlinear Fourier domain. Evidences are found that the eigenvalues and the nonlinear spectral amplitudes are effected by the noise. And the fluctuations in the eigenvalues and nonlinear spectral amplitudes increase with the increase of noise. However, the absolute change in the eigenvalues and nonlinear spectral amplitudes is negligible and will not influence and implementation of the nonlinear frequency division multiplexing.

\end{document}